\documentclass[submission, Proceedings]{SciPost}

\hypersetup{
    colorlinks,
    linkcolor={red!50!black},
    citecolor={blue!50!black},
    urlcolor={blue!80!black}
}

\usepackage[bitstream-charter]{mathdesign}
\DeclareSymbolFont{usualmathcal}{OMS}{cmsy}{m}{n}
\DeclareSymbolFontAlphabet{\mathcal}{usualmathcal}

\begin{document}

% TODO: write your article's title here.
% The article title is centered, Large boldface, and should fit in two lines
\begin{center}{\Large \textbf{
The LHCspin project\\
}}\end{center}

% TODO: write the author list here. Use initials + surname format.
% Separate subsequent authors by a comma, omit comma at the end of the list.
% Mark the corresponding author with a superscript *.
\begin{center}
M. Santimaria\textsuperscript{1$\star$},
V. Carassiti\textsuperscript{2},
G. Ciullo\textsuperscript{2,3},
P. Di Nezza\textsuperscript{1},
P. Lenisa\textsuperscript{2,3},
S. Mariani\textsuperscript{4,5},
L. L. Pappalardo\textsuperscript{2,3} and
E. Steffens\textsuperscript{6}
\end{center}

% TODO: write all affiliations here.
% Format: institute, city, country
\begin{center}
{\bf 1} INFN Laboratori Nazionali di Frascati, Frascati, Italy
\\
{\bf 2} INFN Ferrara, Italy
\\
{\bf 3} Department of Physics, University of Ferrara, Italy
\\
{\bf 4} INFN Firenze, Italy
\\
{\bf 5} Department of Physics, University of Firenze, Italy
\\
{\bf 6} Physics Dept., FAU Erlangen-Nurnberg, Erlangen, Germany
\\
% TODO: provide email address of corresponding author
* marco.santimaria@lnf.infn.it
\end{center}

\begin{center}
\today
\end{center}

% For convenience during refereeing (optional),
% you can turn on line numbers by uncommenting the next line:
%\linenumbers
% You should run LaTeX twice in order for the line numbers to appear.

\definecolor{palegray}{gray}{0.95}
\begin{center}
\colorbox{palegray}{
  \begin{tabular}{rr}
  \begin{minipage}{0.1\textwidth}
    \includegraphics[width=22mm]{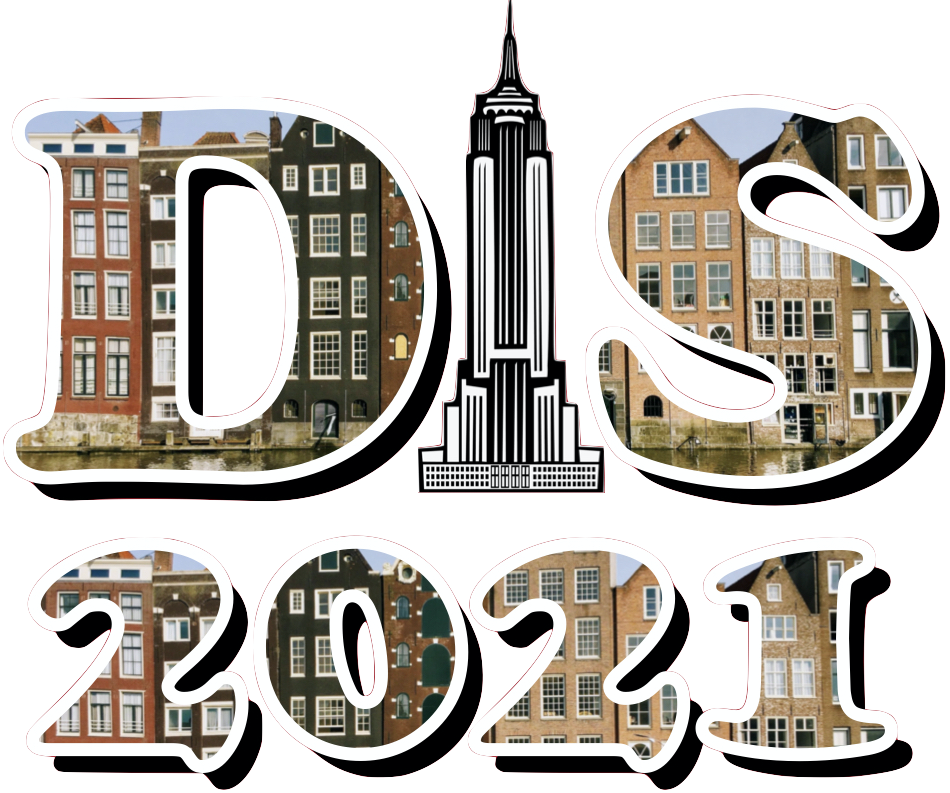}
  \end{minipage}
  &
  \begin{minipage}{0.75\textwidth}
    \begin{center}
    {\it Proceedings for the XXVIII International Workshop\\ on Deep-Inelastic Scattering and
Related Subjects,}\\
    {\it Stony Brook University, New York, USA, 12-16 April 2021} \\
    \doi{10.21468/SciPostPhysProc.?}\\
    \end{center}
  \end{minipage}
\end{tabular}
}
\end{center}

\section*{Abstract}
{\bf
% TODO: write your abstract here.
Broad and unexplored kinematic regions can be accessed at the LHC with fixed-target $pp$, $pA$ and $PbA$ collisions at $\sqrt{s_{\rm{NN}}}=72-115~\rm{GeV}$. 
The LHCb detector is a fully-instrumented forward spectrometer able to run in fixed-target mode, and currently hosts a target gas cell to take data in the upcoming Run 3.
The LHCspin project aims at extending this physics program to Run 4 and to bring polarised physics at the LHC.
An overview of the physics potential and a description of the LHCspin experimental setup are presented.

%The abstract is in boldface, and should fit in 8 lines.
%It should be written in a clear and accessible style, emphasizing %the context, the problem(s) studied, the methods used, the results %obtained, the conclusions reached, and the outlook. You can add a %table contents, recommended if your paper is more than 6 pages long.
}

% TODO: include a table of contents (optional)
% Guideline: if your paper is longer that 6 pages, include a TOC
% To remove the TOC, simply cut the following block
%\vspace{10pt}
%\noindent\rule{\textwidth}{1pt}
%\tableofcontents\thispagestyle{fancy}
%\noindent\rule{\textwidth}{1pt}
%\vspace{10pt}

\section{Introduction}
\label{sec:intro}
% TODO: write your article here.
The LHC delivers proton and lead beams\footnote{A short run with xenon ions was performed in 2017, while an oxygen beam is foreseen for Run 3~\cite{Citron:2018lsq}.}with an energy of $7~\rm{TeV}$ and $2.76~\rm{TeV}$ per nucleon, respectively, with world's highest intensity.
Fixed-target collisions, occurring at an energy in the center of mass of up to $115~\rm{GeV}$, thus offer an unprecedented opportunity to investigate partons carrying a large fraction of the target nucleon momentum.

\begin{figure}[h]
\centering
\includegraphics[width=0.49\textwidth]{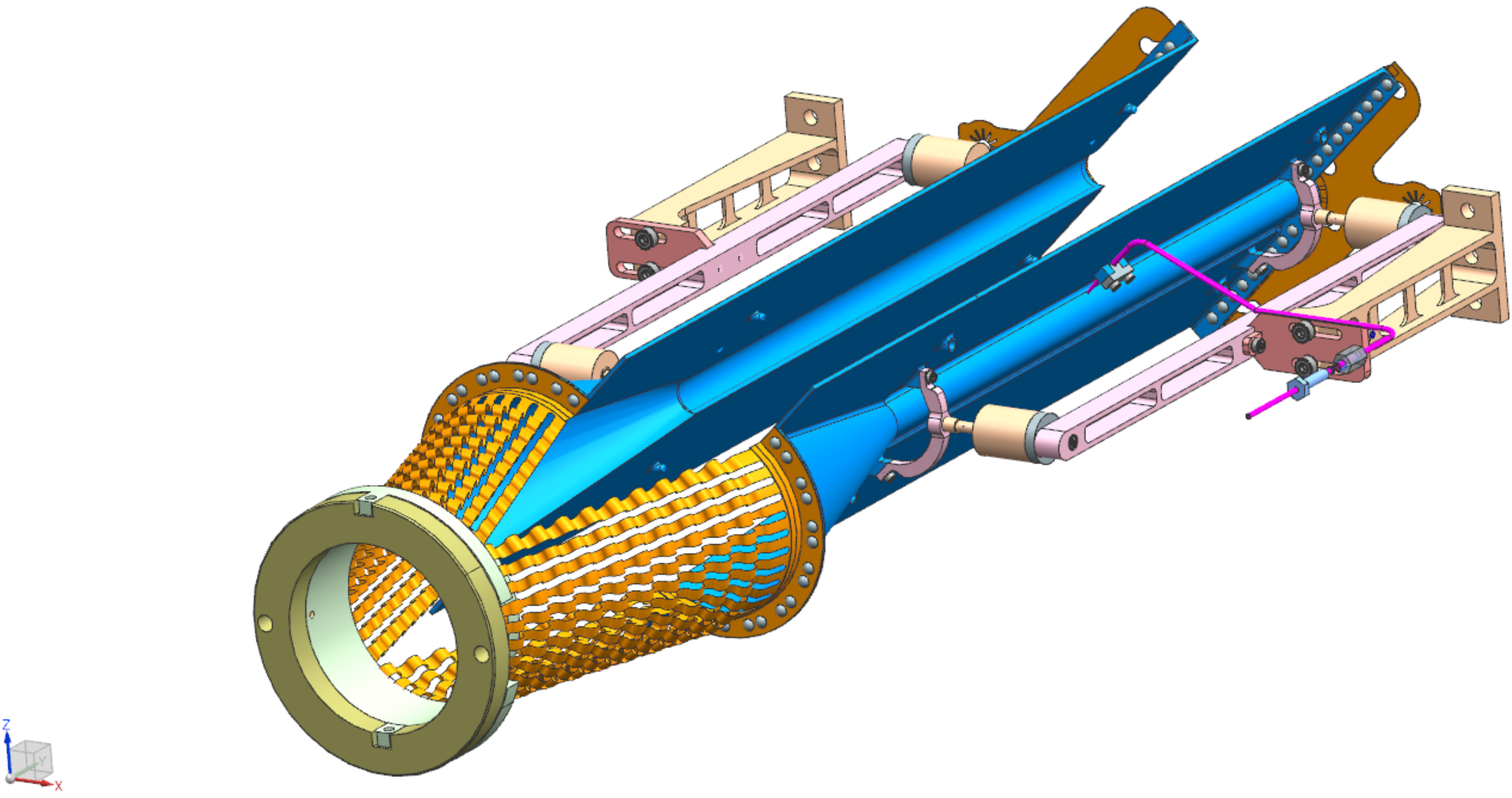}
\includegraphics[width=0.49\textwidth]{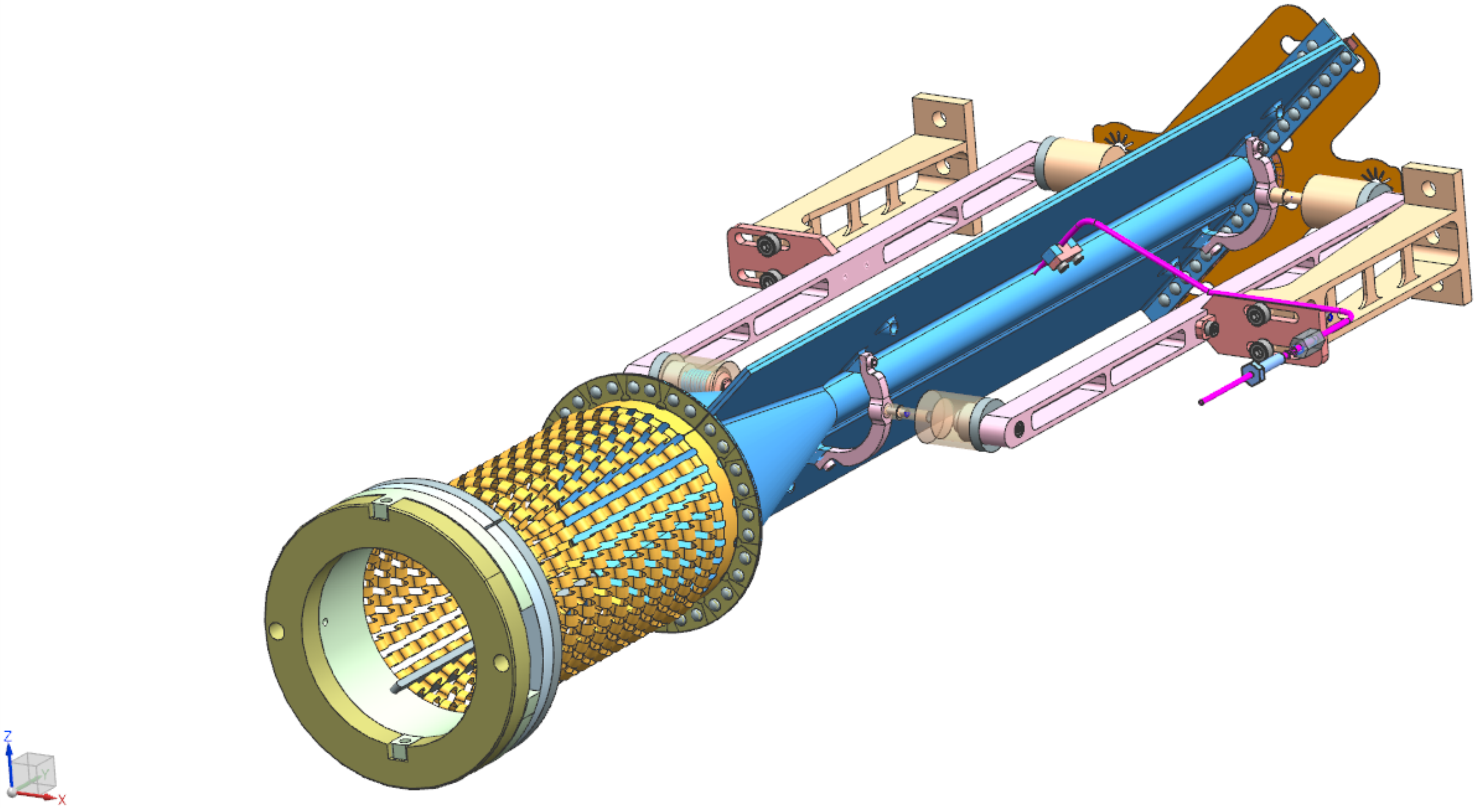}
\caption{SMOG2 storage cell in open (left) and closed (right) configuration.}
\label{fig:smog2}
\end{figure}

The LHCb detector~\cite{Alves:2008zz} is a general-purpose forward spectrometer specialised in detecting hadrons containing $c$ and $b$ quarks, and the only LHC detector able to collect data in both collider and fixed-target mode.
The fixed-target physics program at LHCb is active since the installation in Run 2 of the SMOG (System for Measuring the Overlap with Gas) device~\cite{Aaij:2014ida}, enabling the injection of noble gases in the beam pipe section crossing the VELO (Vertex Locator) detector at a pressure of $\mathcal{O}(10^{-7})~\rm{mbar}$.
With the SMOG2 upgrade~\cite{LHCbCollaboration:2673690}, an openable gas storage cell, shown in Fig.~\ref{fig:smog2}, has been installed in 2020 in front of the VELO. This boosts the target areal density by a factor of $8$ to $35$ depending on the injected gas species. SMOG2 data will be collected in the upcoming Run 3 with a novel reconstruction software allowing simultaneous data-taking of beam-gas and beam-beam collisions, as shown in Fig.~\ref{fig:kin}.

The LHCspin project~\cite{Aidala:2019pit} aims at extending the SMOG2 physics program in Run 4 and, with the installation of a polarised gas target, to bring spin physics at LHC by exploiting the well suited LHCb detector. 
A selection of physics opportunities accessible at LHCspin is presented in Sec.~\ref{sec:phys}, while the experimental setup is discussed in Sec.~\ref{sec:det}.

\section{Physics case}
\label{sec:phys}
The physics case of LHCspin covers three main areas: exploration of the wide physics potential offered by unpolarised gas targets, investigation of the nucleon spin and heavy ion collisions.

\begin{figure}[h]
\centering
\includegraphics[width=0.52\textwidth]{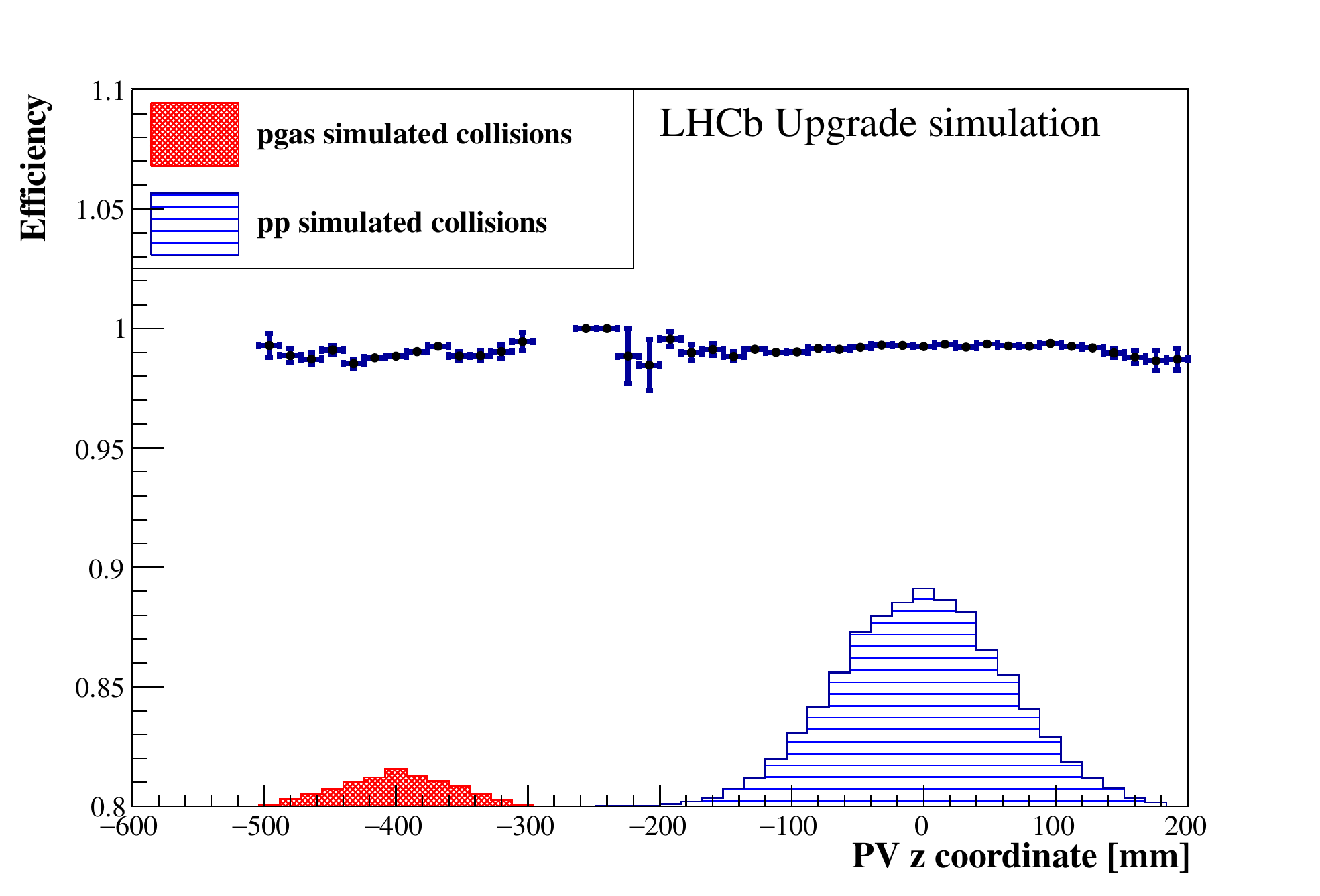}
\hfill
\includegraphics[width=0.47\textwidth]{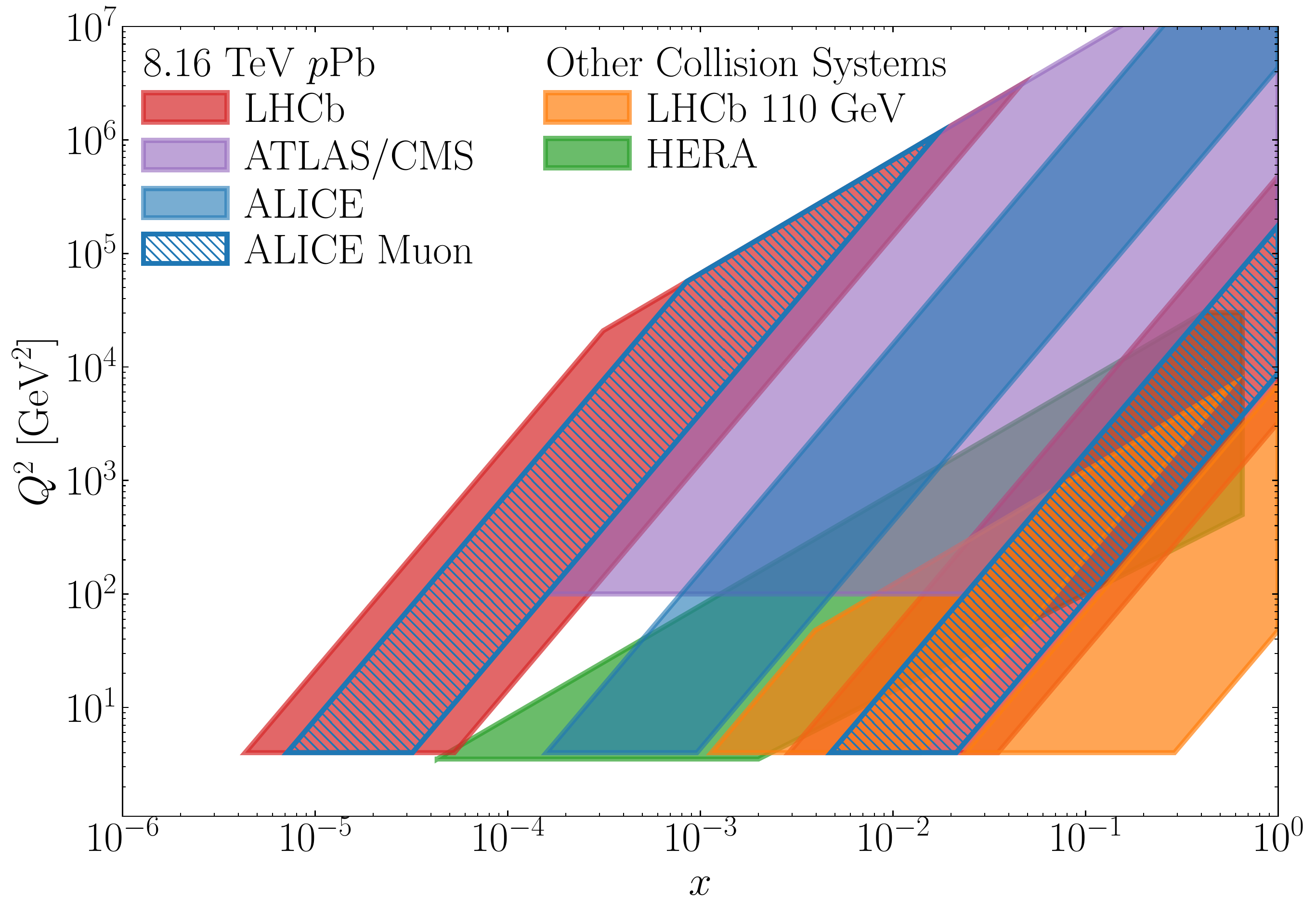}
\caption{Left: Reconstruction efficiency for beam-gas (red) and beam-beam (blue) primary vertices~\cite{LHCB-FIGURE-2019-007}. Right: kinematic coverage of LHCspin (orange) and other existing facilities.}
\label{fig:kin}
\end{figure}

\paragraph{Unpolarised gas targets}
Similarly to SMOG2, LHCspin will allow the injection of several species of unpolarised gases: $H_2$, $D_2$, $He$, $N_2$, $O_2$, $Ne$, $Ar$, $Kr$ and $Xe$ with negligible impact on the LHC beam lifetime. This gives an excellent opportunity to investigate parton distribution functions (PDFs) in both nucleons and nuclei in the large-$x$ and intermediate $Q^2$ regime (Fig.~\ref{fig:kin}), which is especially affected by lack of experimental data and impact several fields from basic QCD tests to astrophysics.
For example, the large acceptance and high reconstruction efficiency of LHCb on heavy flavour enables the study of gluon PDFs, which are a fundamental input for theoretical predictions~\cite{Hadjidakis:2018ifr}, while measurements of heavy-flavour hadroproduction directly impact prompt muonic neutrino flux knowledge~\cite{Garzelli:2016xmx}.
Searches for an intrinsic charm component in the proton~\cite{Aaij:2018ogq} and prompt antiproton production in $pHe$ collisions~\cite{Aaij:2018svt} are two other high-profile example measurements that have already been pioneered at LHCb. 
With the large amount of data to be collected with LHCspin, nuclear PDFs can also be investigated in greater detail, helping to shed light on the intriguing anti-shadowing effect~\cite{Eskola:2016oht}.

\begin{figure}[h]
\centering
\includegraphics[width=0.49\textwidth]{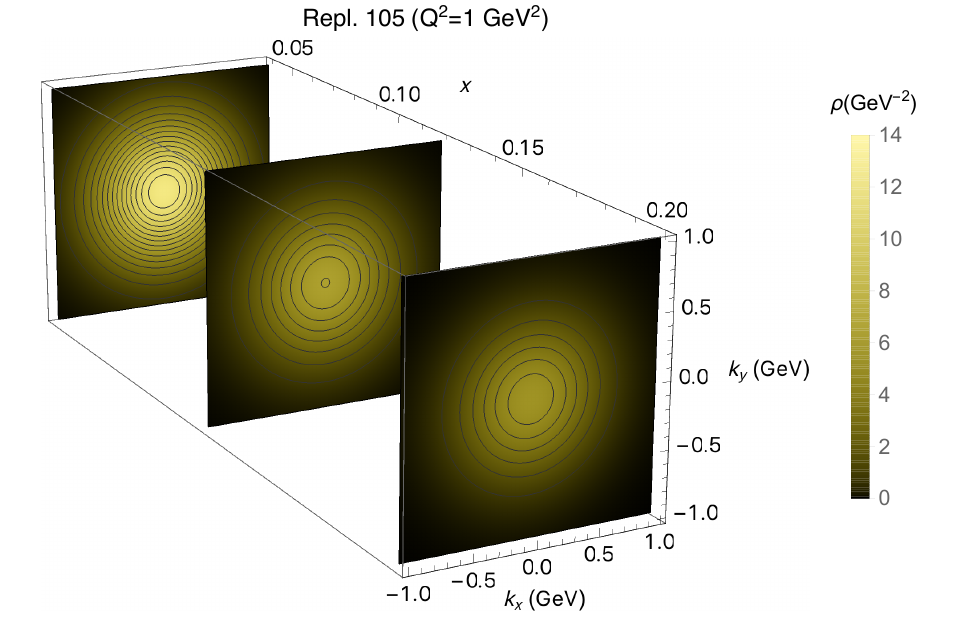}
\hfill
\includegraphics[width=0.35\textwidth]{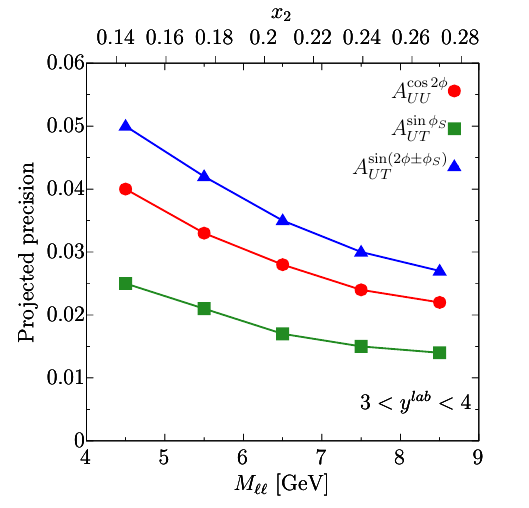}
\caption{Left: up quark densities in momentum space~\cite{Bacchetta:2017gcc}. Right: projected precision for some TSSAs with Drell-Yan data~\cite{Hadjidakis:2018ifr} with $x_2$ being the longitudinal momentum fraction of the target nucleon and $\rm{M}_{\ell\ell}$ the dilepton invariant mass.}
\label{fig:pdfs}
\end{figure}

\paragraph{Spin physics}
Beside standard collinear PDFs, LHCspin will offer the opportunity to probe polarised quark and gluon distributions by means of proton collisions on polarised hydrogen and deuterium.
For example, measurements of transverse-momentum dependent PDFs (TMDs) provide a map of parton densities in 3-dimensional momentum space, as sketched in Fig.~\ref{fig:pdfs}.
Light quark TMDs, especially in the high-$x$ regime, can be accessed by measuring transverse single spin asymmetries (TSSAs) in Drell-Yan or weak-boson production processes, while gluon densities, such as the gluon Sivers function, can be probed via heavy-flavour production.
Fig.~\ref{fig:pdfs} shows the projected precision for some TSSAs at LHCspin with $10~\rm{fb}^{-1}$ of data.

It is also attractive to go beyond a 3-dimensional description by building observables which are sensitive to Wigner distributions~\cite{Bhattacharya:2017bvs} and to measure the elusive
transversity PDF, whose knowledge is currently limited to valence quarks at the leading order~\cite{Radici:2018iag}, as well as its integral, the tensor charge, which is of direct interest in constraining physics beyond the Standard Model~\cite{Courtoy:2015haa}. 

\paragraph{Heavy ion collisions}
Thermal heavy-flavour production is negligible at the typical temperature of few hundreds MeV of the system created in heavy-ion collisions. Quarkonia states ($c\overline{c}$, $b\overline{b}$) are instead produced on shorter timescales, and their energy change while traversing the medium represents a powerful way to investigate Quark-Gluon Plasma (QGP) properties. LHCb capabilities allow to both cover the aforementioned charmonia and bottomonia studies and to extend them to bottom baryons as well as exotic probes.  
QGP phase diagram exploration at LHCspin can be performed with a rapidity scan, complementing RHIC's beam-energy scan, while flow measurements will greatly benefit from the excellent LHCb identification performance on charged and neutral light hadrons.
An interesting topic joining heavy-ion collisions and spin physics is the dynamics of small systems which can be probed via ellipticity measurement in lead collisions on polarised deuterons~\cite{Broniowski:2019kjo}.

\section{Experimental setup}
\label{sec:det}
\begin{figure}[h]
\centering
\includegraphics[width=0.6\textwidth]{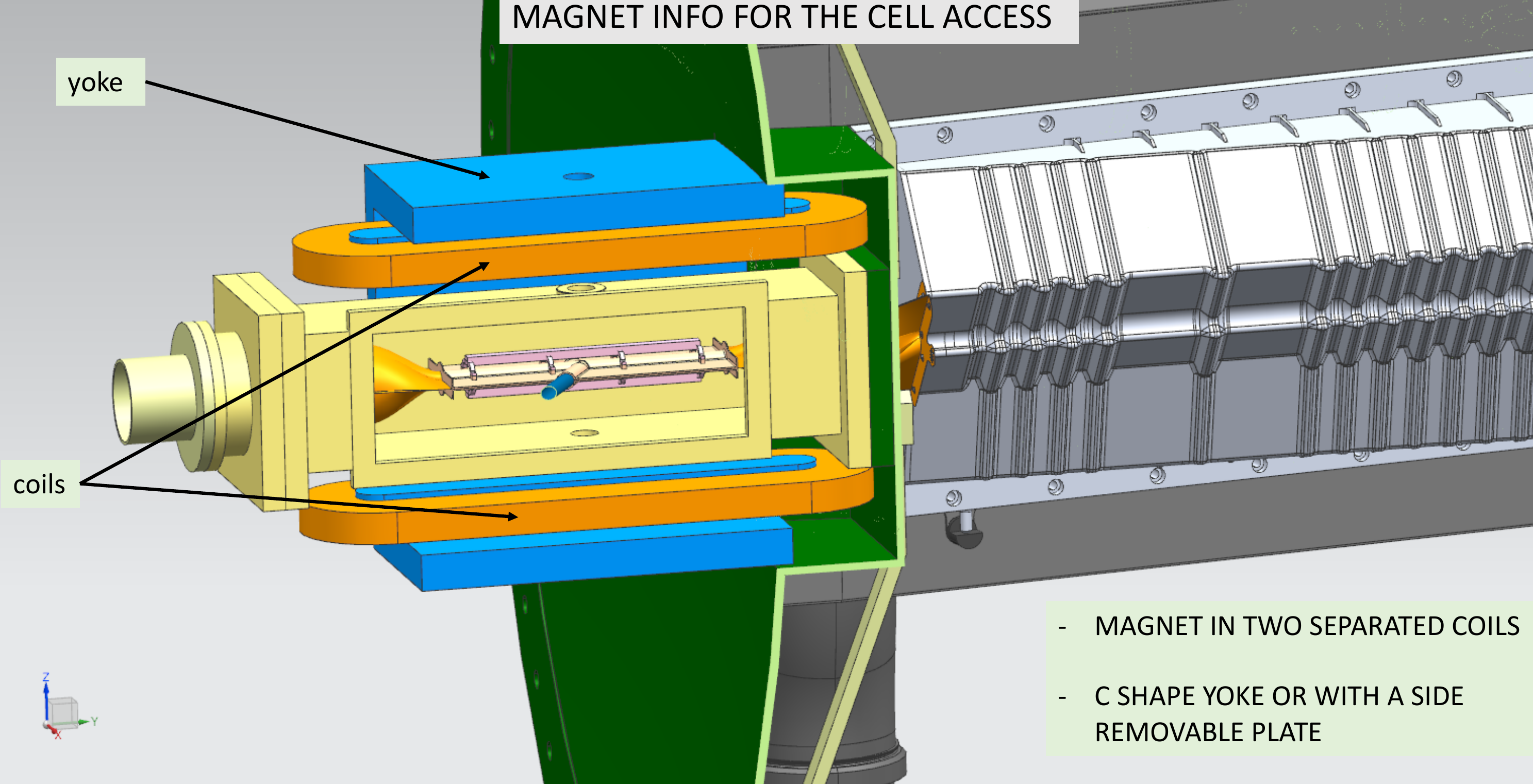}
\caption{The TC drawing with the magnet coils (orange) and the iron return yoke (blue) enclosing the storage cell. The VELO vessel and detector box are shown in green and grey, respectively.}
\label{fig:rd}
\end{figure}

The LHCspin experimental setup is in R\&D phase and calls for the development of a new generation polarised target. The starting point is the setup of the HERMES experiment at DESY~\cite{Airapetian:2004yf} and comprises three main components: an Atomic Beam Source (ABS), a Target Chamber (TC) and a diagnostic system.
The ABS consists of a dissociator with a cooled nozzle, a Stern-Gerlach apparatus to focus the wanted hyperfine states, and adiabatic RF-transitions for setting and switching the target polarisation between states of opposite sign.
The ABS injects a beam of polarised $H$ or $D$ into the TC, which is located into the LHC primary vacuum.
The TC hosts a T-shaped openable storage cell, sharing the SMOG2 design, and a dipole holding magnet, as shown in Fig.~\ref{fig:rd}.

The diagnostic system continuously analyses gas samples drawn from the TC and comprises a target gas analyser to detect the molecular fraction, and thus the degree of dissociation, and a Breit-Rabi polarimeter to measure the relative population of the injected hyperfine states.
An instantaneous luminosity of $\mathcal{O}(10^{32})~\rm{cm}^{-2}\rm{s}^{-1}$ is foreseen for $pH$ collisions at Run 4.

\section{Conclusion}
The fixed-target physics program at LHC has been greatly enhanced with the recent installation of the SMOG2 gas storage cell at LHCb.
LHCspin is the natural evolution of SMOG2 and aims at installing a polarised gas target to bring spin physics at LHC for the first time, opening a whole new range of exploration.
With strong interest and support from the international theoretical community, LHCspin is a unique opportunity to advance our knowledge on several unexplored QCD areas, complementing both existing facilities and the future Eletron-Ion Collider~\cite{Accardi:2012qut}.

% TODO: include funding information
\paragraph{Funding information}
The project leading to this application has received funding from the INFN (Italy) and the European Union's Horizon 2020 research and innovation program. %Correctly-provided data will be linked to funders listed in the \href{https://www.crossref.org/services/funder-registry/}{\sf Fundref registry}.

\bibliography{main.bib}

\nolinenumbers

\end{document}